\documentclass[aps,prl,amsmath,amssymb,floatfix,twocolumn,showpacs,10pt]{revtex4-1}

\usepackage{times}
\usepackage{bbold} 
\usepackage{bm}
\usepackage{graphicx} 
\usepackage{color}
\usepackage{hyperref}
\usepackage{textcomp}
\usepackage{amssymb} 
\usepackage{wasysym}
\usepackage{mathptmx}  
\usepackage{multibib} 
\newcites{supp}{Supplemental References}

\begin{document}

\title{Nuclear-spin-induced localization of the edge states in two-dimensional topological insulators}

\author{Chen-Hsuan Hsu$^{1}$}
\author{Peter Stano$^{1,2}$}
\author{Jelena Klinovaja$^{1,3}$}
\author{Daniel Loss$^{1,3}$}

\affiliation{$^{1}$RIKEN Center for Emergent Matter Science (CEMS), Wako, Saitama 351-0198, Japan}
\affiliation{$^{2}$Institute of Physics, Slovak Academy of Sciences, 845 11 Bratislava, Slovakia}
\affiliation{$^{3}$Department of Physics, University of Basel, Klingelbergstrasse 82, CH-4056 Basel, Switzerland}

\date{\today}

\begin{abstract}

We investigate the influence of nuclear spins on the resistance of helical edge states of two-dimensional topological insulators (2DTIs). Via the hyperfine interaction, nuclear spins allow electron backscattering, otherwise forbidden by time-reversal symmetry. 
We identify two backscattering mechanisms, depending on whether the nuclear spins are ordered or not. Their temperature dependence is distinct but both give resistance, which increases with the edge length, decreasing temperature, and increasing strength of the electron-electron interaction.
Overall, we find that the nuclear spins will typically shut down the conductance of the 2DTI edges at zero temperature.
 
\end{abstract}


\maketitle

Two-dimensional topological insulators (2DTIs), such as HgTe/(Hg,Cd)Te~\cite{Bernevig:2006,Konig:2007} and InAs/GaSb quantum wells~\cite{Liu:2008,Knez:2011}, have potential in dissipationless transport and quantum computation~\cite{Hasan:2010,Qi:2011}. 
The hallmark of 2DTIs is helical states propagating along the edges. Since the elastic edge electron backscattering requires a spin flip, the edge channel conductance is immune against time-reversal invariant perturbations, covering dominant disorder forms. Experiments, however, did not show robustly quantized conductance~\cite{Konig:2007,Roth:2009,Knez:2011,Charpentier:2013,Du:2015,Nichele:2016}, which initiated extensive investigations on possible backscattering mechanisms. Various sources of resistance were proposed, such as single~\cite{Wu:2006,Maciejko:2009,Tanaka:2011,Vayrynen:2016} and a bath of~\cite{Hattori:2011,Lunde:2012,Altshuler:2013,Vayrynen:2016} magnetic impurities, random magnetic fluxes~\cite{Delplace:2012}, random Rashba spin-orbit coupling in the presence of an Overhauser field~\cite{DelMaestro:2013} or inelastic scattering~\cite{Strom:2010,Crepin:2012,Geissler:2014}, phonons~\cite{Budich:2012}, multi-particle scattering~\cite{Lezmy:2012,Schmidt:2012,Kainaris:2014}, or coupling to disorder-localized states with spin~\cite{Vayrynen:2013}.
 
Here we identify nuclear spins as an omnipresent source of resistance for 2DTI edge channels. At first sight, this might come as a surprise given that the strength of the hyperfine interaction between nuclear spins and itinerant electrons is very weak~\cite{Braun:2006} and for noninteracting electrons results in negligible resistance. However, as is well known, electron-electron interactions strongly amplify the backscattering effects in one-dimensional geometries~\cite{Giamarchi:1988,Giamarchi:2003}. Indeed, we find that if the edge channels are long and the electron-electron interactions are strong, nuclear spins generally are a relevant resistance source at dilution fridge temperatures. For typical experimental conditions, the hyperfine-induced backscattering can be amplified even up to the strong-coupling regime, resulting in an exponentially small edge conductance. 

The physics beyond this simple observation gets complicated by the fact that nuclear spins can order under certain conditions such as low temperatures and strong interactions~\cite{Simon:2007,Simon:2008,Braunecker:2009a,Braunecker:2009b,Scheller:2014,Meng:2014a}. This ordering is a result of the Ruderman-Kittel-Kasuya-Yosida (RKKY) interaction between the nuclear spins, mediated by the itinerant edge electrons.
On one hand, the ordered nuclear spins become ineffective in backscattering since the electron-nuclear spin flip-flop requires now an energy (to emit a magnon) much larger than the temperature. Ordering therefore screens nuclear spins (and possibly additional magnetic impurities) from backscattering electrons, and the resistance should decrease upon lowering the temperature. On the other hand, nuclear spin ordering produces a macroscopic magnetic (Overhauser) field which breaks the time-reversal symmetry. This field allows for backscattering on ordinary static potential disorder (referred to as `impurities' henceforth, not to be confused with the (dis-)order in the nuclear spin orientation), and the associated resistance increases upon lowering the temperature. Finally, because the RKKY interaction between the nuclear spins is mediated by edge electrons, the two subsystems enter a complex interdependence, giving rise to a rich behavior of the edge resistance as a function of temperature.

Here we determine this temperature behavior by performing renormalization-group (RG) analysis for the electron-nuclear system in the presence of interactions and impurities, both above and below the expected ordering temperature. We find that for relevant parameter values the most typical scenario is as follows. At few Kelvins, the nuclear spins are thermally disordered and induce resistance with a power-law temperature dependence, which, for sufficiently long edges, evolves into an exponential well below 1 Kelvin. For strongly interacting (say, the Luttinger liquid parameter $K=0.2$) and long edges (the edge length $L$ of the order of tens of $\mu$m), this resistance can be of the order of the quantum resistance. Once the nuclear spins order (a typical ordering temperature $T_0$ is of the order of tens of mK), they establish a finite Overhauser field, which allows backscattering on impurities and results in an exponentially growing resistance. The characteristic temperature dependence of this exponential, markedly different from the case of a nonhelical, spin-degenerate channel, would be an indication of both the nuclear spin ordering as well as {\it the helical nature of the edge channel itself}. 

\begin{figure}[t]
\centering
\includegraphics[width=\linewidth]{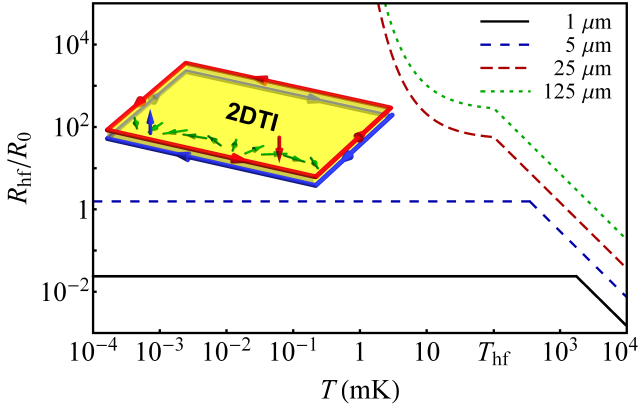}
\caption{
Temperature ($T$) dependence of the resistance induced by thermally disordered nuclei for various edge lengths $L$. The localization-delocalization transition (from a power law to an exponential) is visible when $L>\xi_{\textrm{hf}}$ (the two topmost curves).
Inset: 2DTI helical edges with up-spin (blue) and the down-spin (red) electrons moving in opposite directions (routes are separated for clarity). The spin quantization ($z$) axis is perpendicular to the 2DTI plane. The nuclear spins at the boundaries (green arrows) are ordered \cite{Braunecker:2009a,Braunecker:2009b} below the transition temperature $T_{0}$, and become randomly oriented (not shown) above it. For clarity, spins are drawn only at one edge.
}
\label{Fig:Rhf_T}
\end{figure}

{\it Hamiltonian and backscattering action.} We model the edge electrons and the nuclear spins (see the inset of Fig.~\ref{Fig:Rhf_T}) with the Hamiltonian, $\mathit{H}=\mathit{H}_{\textrm{el}}+\mathit{H}_{\textrm{hf}}$. 
The electrons are described as a helical Tomonaga-Luttinger liquid,
\begin{equation}
\mathit{H}_{\textrm{el}} = \int \frac{\hbar dr}{2\pi} \, \left\{ uK \left[ \partial_{r} \theta(r) \right]^2 + \frac{u}{K} \left[ \partial_{r} \phi(r) \right]^2 \right\},
\label{eq:H_LL}
\end{equation}
where $\theta$ and $\phi$ are bosonic fields, functions of the edge coordinate $r$, parametrizing the left-moving up-spin $L_{\uparrow}$ and right-moving down-spin $R_{\downarrow}$ fermionic fields. The parameter $K$ relates the renormalized velocity $u=v_{F}/K$ to the Fermi velocity $v_{F}$ (with the Fermi energy $\epsilon_{F} \equiv \hbar v_{F} k_{F}/2$ and the Fermi wave vector $k_F$). The bosonization requires a short-distance cutoff, taken as $a=\hbar v_{F}/\Delta$, the transverse decay length of the edge electron wave function defined by $\Delta$, the 2DTI bulk gap.

The hyperfine interaction,
\begin{equation}
\mathit{H}_{\textrm{hf}} = \frac{A_{0}}{\rho_{\textrm{nuc}}}  \sum_{n} \delta({\bf x}-{\bf x}_{n}) \frac{\boldsymbol{\sigma}}{2} \cdot {\bf I}_{n},
\label{Eq:H_hf}
\end{equation}
describes the coupling of the electron spin $\boldsymbol{\sigma}/2$ to nuclear spins ${\bf I}_n$ at positions ${\bf x}_n$ labeled by index $n$. Here $A_{0}$ is the hyperfine coupling, and $\rho_{\textrm{nuc}} = 8/a_{0}^{3}$ is the nuclear density with the lattice constant $a_{0}$. For simplicity, we assume a homonuclear system, and neglect the variation of the edge electron wave function in the transverse direction such that it is given by $1 / \sqrt{W a}$, with the quantum well thickness $W$. This reduces the problem dimensionality, as now electrons interact with effective spins of the whole cross section, a sum of $N_\perp$ nuclear spins (each with magnitude $I$). In Eq.~(\ref{Eq:H_hf}) we take the Fermi contact hyperfine interaction, with dipole-dipole and orbital contributions~\cite{Okvatovity:2016} much weaker (see Supplemental Material (SM) for a comparison~\cite{SM}). Whereas the dipole-dipole interaction between the nuclear spins is not considered in Eqs.~(\ref{eq:H_LL})--(\ref{Eq:H_hf}), we include it in our analysis as the spin dissipation mechanism for the nuclei~\cite{SM,Kornich:2015}.

Unless stated otherwise, we adopt parameters of InAs/GaSb, namely $v_{F}=4.6 \times 10^4~$m/s~\cite{Pribiag:2015,Schrade:2015}, $a_{0}=6.1~$\AA, $\Delta=3.4~$meV, $a=9$~nm~\cite{Li:2015}, $W=20~$nm~\cite{Li:2015,Nichele:2016}, $K=0.2$ (the reported values vary from $0.2$ to $0.9$~\cite{Wu:2006,Maciejko:2009,Hou:2009,Strom:2009,Teo:2009,Egger:2010,Li:2015}), $k_{F}=7.9 \times 10^{7}~$m$^{-1}$~\cite{Knez:2011}, $A_{0}=50~\mu$eV~\cite{Gueron:1964,Lunde:2013,Paget:1977,Schliemann:2003,Braun:2006}, $I=3$ (the approximate average of all constituent isotopes), and $N_{\perp}=3900.$

We derive the nuclear spin contribution to the electronic imaginary-time action as
\begin{eqnarray}  
\frac{ \delta \mathit{S}}{\hbar} &=& 
- D \int_{u|\tau-\tau'|>a} \frac{v_{F}^2 dr d\tau d\tau'}{8\pi a^3} \; e^{-\omega|\tau-\tau'|}
\nonumber \\
&& \hspace{0.85in} \times 
\cos \left[ 2 \phi (r,\tau)- 2\phi (r,\tau') \right], 
\label{Eq:S_bs}
\end{eqnarray}
with $D$ a prefactor and $\hbar \omega$ the energy cost of nuclear spin flip accompanying the electron backscattering. We specify these two factors for various mechanisms below, and analyze the resistance building the RG equations~\cite{Giamarchi:1988,Giamarchi:2003} based on Eqs.~\eqref{eq:H_LL} and \eqref{Eq:S_bs}.

{\it Elastic backscattering on disordered nuclear spins.}
We first consider thermally disordered nuclear spins (i.e., randomly oriented, including those within a cross section), which is the most typical situation. Averaging over such random spins, we get $D_{\textrm{hf}} = A_{0}^2 I(I+1)/(3 \pi N_{\perp} \Delta^2)$, and, since they can be flipped at no cost, $\omega_{\textrm{hf}}=0$. We note that the backscattering becomes stronger upon decreasing $N_{\perp}$, and 
 is RG relevant for $K<3/2$, so that electrons with repulsive interactions ($K<1$) get localized. The resistance of an edge longer than the associated localization length 
$\xi_{\textrm{hf}} = a (K^2 D_{\textrm{hf}})^{-1/(3-2K)} $ grows exponentially below the localization temperature $T_{\textrm{hf}}\equiv \hbar u / (k_{B} \xi_{\textrm{hf}})$. For our parameters, $\xi_{\textrm{hf}} \approx 17~\mu\textrm{m}$ and $T_{\textrm{hf}}\approx 100~$mK give scales at which this resistance source becomes important. 
It shows that backscattering by thermally disordered nuclear spins can strongly affect edge states.

We now proceed to explicit formulas. At zero bias, we identify three regimes, depending on which is the shortest among the thermal length $\lambda_{T}\equiv \hbar u / (k_{B} T)$, the localization length $\xi_{\textrm{hf}}$, and the edge length $L$. First, for $\lambda_{T}<L,~\xi_{\textrm{hf}}$, we get
\begin{eqnarray}
R_{\textrm{hf}} (T) & \propto & R_{0} \frac{\pi D_{\textrm{hf}} L }{2 a} \left( \frac{K k_{B}T}{\Delta} \right)^{2K-2},
\label{Eq:R_hf1}
\end{eqnarray} 
with $R_{0} \equiv h/e^2$.
Second, if $\xi_{\textrm{hf}} < \lambda_{T},~L$, the edge is gapped, with a thermally activated resistance,
\begin{eqnarray}
R_{\textrm{hf}} (T) &\propto& R_{0} \frac{\pi D_{\textrm{hf}}L }{2 a}  e^{\Delta_{\textrm{hf}} /(k_{B}T)},
\label{Eq:R_hf2}
\end{eqnarray} 
and the gap $\Delta_{\textrm{hf}} = \Delta \left( 2K^3 D_{\textrm{hf}} \right)^{1/(3-2K)}\approx 1.2~\mu$eV.   
Finally, if $L< \xi_{\textrm{hf}},~\lambda_{T}$, we obtain
\begin{eqnarray}
R_{\textrm{hf}} (L) &\propto& R_{0} \frac{\pi D_{\textrm{hf}}L }{2 a} \left( \frac{L}{a} \right)^{2-2K}.
\label{Eq:R_hf3}
\end{eqnarray}
Here we give the resistance $R$ of the helical Tomonaga-Luttinger liquid. Other resistances possibly contribute, in series, to the total edge resistance $R_{\rm tot}$. Most notable is the contact resistance, equal to $R_0$ for a single channel wire. Note that we discuss $R$, not $R_{\rm tot}$, throughout this article. The resistance given by Eqs.~(\ref{Eq:R_hf1})--(\ref{Eq:R_hf3}) is plotted in Fig.~\ref{Fig:Rhf_T}, as a function of the temperature. Upon decreasing the temperature $T$ from few Kelvins, the resistance first increases as a power law, and then saturates (for short edges) or grows exponentially (for long edges).

\begin{figure}[t]
\centering
\includegraphics[width=\linewidth]{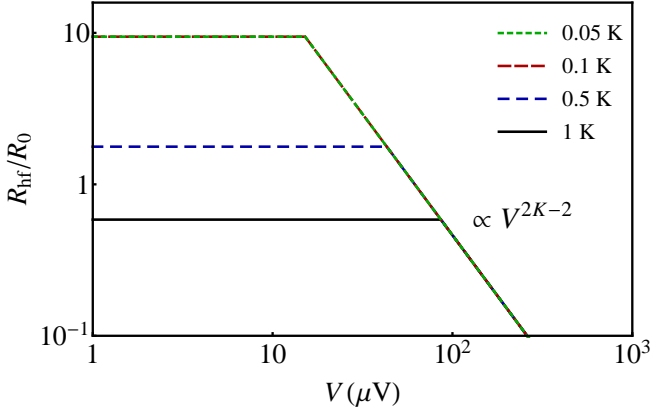}
\caption{
Bias voltage ($V$) dependence of the differential resistance for $L=10~\mu$m and disordered nuclei for various temperatures. 
}
\label{Fig:Rhf_V}
\end{figure}

Let us now consider a finite bias voltage $V$, plotting the differential resistance of an edge shorter than $\xi_{\textrm{hf}}$ in Fig.~\ref{Fig:Rhf_V}. At high bias, $\lambda_{V} \equiv \hbar u / (eV) < L,~\lambda_{T}$, the differential resistance is given by Eq.~(\ref{Eq:R_hf1}) upon the replacement $\lambda_{T} \rightarrow \lambda_{V}$. It grows with a decreasing voltage as a power law, before it saturates at a value determined by the shorter of $\lambda_{T}$ and $L$, Eqs.~\eqref{Eq:R_hf1} and \eqref{Eq:R_hf3}, respectively. A fractional power-law dependence of the edge conductance on both the temperature and the bias voltage has been observed in InAs/GaSb 2DTIs with short edges~\cite{Li:2015}, though not attributed to nuclear spins.

{\it Nuclear spin order.} 
We now consider the scenario in which nuclear spins are ordered. The ordering, predicted to occur generally in quasi one-dimensional finite-size conductors~\cite{Simon:2008,Braunecker:2009a,Braunecker:2009b,Klinovaja:2013b,Meng:2014a,Hsu:2015}, is stabilized by the RKKY interaction mediated by edge electrons~\cite{SM}. This interaction results in nuclear spins aligning ferromagnetically within a cross section, along a vector which rotates in space upon moving along the edge with a period $\pi/k_F$. 
Performing the spin-wave analysis along the line of Refs.~\cite{Braunecker:2009b,Meng:2014a,Hsu:2015}, we find that the transition temperature is higher for a helical conductor ($T_{0}\approx 42~$mK for our parameters) than a spin-degenerate wire, indicating that the system tendency toward ordering is higher for a helical conductor. Further, whereas nuclear ordering in a spin-degenerate wire leads to a partial gap~\cite{Scheller:2014,Aseev:2017}, in a helical edge it is energetically favorable {\it not} to open a gap at the Fermi surface~\footnote{However, a gap is induced below the Fermi surface, providing experimental signatures for the ordering~\cite{SM,Dobers:1988,Smet:2002,Chekhovich:2013,Tiemann:2014}.}. 
Nevertheless, the resistance is still influenced by the nuclear ordering, as we now show.

To this end, we write Eq.~\eqref{Eq:H_hf} as a sum, $\mathit{H}_{\textrm{hf}} = \langle \mathit{H}_{\textrm{hf}} \rangle + \mathit{H}_{\textrm{e-mag}}$, of the expectation value in the ordered nuclear state~\cite{Holstein:1940}, 
\begin{eqnarray}
 \langle \mathit{H}_{\textrm{hf}} \rangle &=&   \frac{A_{0}I m_{2k_{F}}}{2\pi a} \int dr \; \cos \left[ 2\phi(r) - 4k_{F} r \right],
\label{Eq:H_Ov}
\end{eqnarray}
being an Overhauser field, and the remainder, being the electron-magnon interaction, 
\begin{eqnarray}
\mathit{H}_{\textrm{e-mag}} &\approx& 
\frac{ A_{0}  }{2  L^2} \sqrt{ \frac{I m_{2k_{F}}}{2 N_{\perp}}} \sum_{q,q'} \; \frac{1}{i} \left( b_{q'}^{\dagger} + b_{- q'} \right)  \nonumber \\
&& \hspace{0.3 in} \times L_{\uparrow}^{\dagger}(q) R_{\downarrow}(q+q'-2k_{F}) + \textrm{H.c.},
\label{Eq:H_e-mag}
\end{eqnarray}
with $b_{q}^\dagger$ creating a magnon with momentum $q$. In the above, $m_{2k_{F}}$ is the order parameter, $m_{2k_{F}}=1$ for completely ordered nuclear spins, and we define the transition temperature by $m_{2k_{F}}(T_{0})=1/2$.
We now analyze the resistance arising from Eqs.~(\ref{Eq:H_Ov}) and (\ref{Eq:H_e-mag}) separately.

{\it Anderson-type localization in the ordered phase.}
Even though the Overhauser field, Eq.~(\ref{Eq:H_Ov}), itself does not lead to backscattering at the Fermi surface, it breaks the time-reversal symmetry and thus lifts the protection of the edge states against impurities. Backscattering can then arise as a second-order process, with the spin flip provided by the Overhauser field and the momentum provided by impurities. We quantify its strength by performing the lowest-order Schrieffer-Wolff transformation~\cite{Schrieffer:1966} and an average over impurities~\cite{Giamarchi:2003}, and obtain Eq.~(\ref{Eq:S_bs}), with $\omega_{\textrm{hx}} = 0$, $ D_{\textrm{hx}} \equiv  D_{b} A_{0}^2 I^2 m_{2k_{F}}^2/(128 \pi a \Delta^2\epsilon_{F}^2)$. 
Inserting numbers, we find that the nuclear order-assisted backscattering on impurities is comparable in strength to backscattering on disordered nuclear spins for an impurity strength $D_b$ corresponding to a bulk mean free path $\lambda_{\textrm{mfp}} \sim 0.1\text{--}1~\mu$m~\cite{Konig:2007,Li:2015}. Because the associated localization temperature $T_{\textrm{hx}} \sim 90\text{--}220~$mK is similar in value to $T_{\textrm{hf}}$, it is typically larger than $T_0$. Equation \eqref{Eq:R_hf2} then applies (with the replacement $\{D_{\textrm{hf}}, \Delta_{\textrm{hf}} \} \to \{D_{\textrm{hx}}, \Delta_{\textrm{hx}} \}$), describing the edge resistance with $\Delta_{\textrm{hx}} = \Delta \left( 2K^3 D_{\textrm{hx}} \right)^{1/(3-2K)}$. The temperature dependence of $\Delta_{\textrm{hx}}$, entering through $m_{2k_F}(T)$, as well as its dependence on $\epsilon_{F}$ and on $D_b$ are the essential differences allowing one to distinguish between the two scenarios, and uncover the nuclear ordering transition.

{\it Magnon-mediated backscattering.}
We finally consider magnons in the nuclear spin system, described by Eq.~\eqref{Eq:H_e-mag}. 
Unlike in the previous cases, the electron spin flip by a magnon now leads to a finite energy exchange. Because our magnons are essentially dispersionless away from zero momentum, we take this energy as momentum independent, $\hbar \omega_{\textrm{mag}} \equiv 2I |J_{2k_{F}}^{x}| m_{2k_{F}}/N_{\perp}$ with the RKKY coupling $J_{2k_{F}}^{x}$~\cite{SM}. 
This approximation allows us to reformulate the magnon-induced backscattering as an effective electron-phonon problem~\cite{Voit:1987}, 
and derive Eq.~(\ref{Eq:S_bs}) with $D_\textrm{mag} = A_{0}^2 I/ (2\pi N_{\perp} \Delta^2)$ and $\omega = \omega_{\textrm{mag}}$. 
From the RG analysis~\cite{Voit:1987,Giamarchi:2003} we are then able to calculate the resistance due to the magnon emission as
\begin{eqnarray}
R_{\textrm{mag}}^{\textrm{em}} (T) &\propto& R_{0} \frac{\pi D_{\textrm{mag}}L}{2 a} \left[ \frac{K \hbar \omega_{\textrm{mag}}(T) }{\Delta} \right]^{2K-3},
\label{Eq:R_mag}
\end{eqnarray} 
which drops with a decreasing temperature as a power law of the magnon energy.
Equation \eqref{Eq:R_mag} is formally valid for $T<T_x$ with $T_x$ defined by $\omega_{\textrm{mag}} (T_{x}) = ( K^2 D_{\textrm{mag}} )^{1/(4-2K)} u/a$, a condition on the validity of the perturbative RG calculation.
We estimate the resistance due to the magnon absorption $R_{\textrm{mag}}^{\textrm{abs}} $ by Eqs.~(\ref{Eq:R_hf1})-(\ref{Eq:R_hf3}), upon the replacement $D_{\textrm{hf}} \to D_{\textrm{hf}} (1-m_{2k_F})$. This essentially means we neglect the magnon energy absorbed by electrons, and consider the contribution from only the disordered nuclear spins, which are present, among all the nuclei, with the weight $(1-m_{2k_F}) \propto T^{3-2K}$. 
We note that, as a consistency check, the total resistance due to magnons, $R_{\textrm{mag}}\equiv R_{\textrm{mag}}^{\textrm{em}}+R_{\textrm{mag}}^{\textrm{abs}}$ should obey a physically motivated upper limit being $R_{\textrm{mag}} {\leq} R_{\textrm{hf}}$, stating that backscatterings penalized by paying an energy cannot lead to a resistance larger than if the energy penalty is removed.

\begin{figure}[t]
\centering
\includegraphics[width=\linewidth]{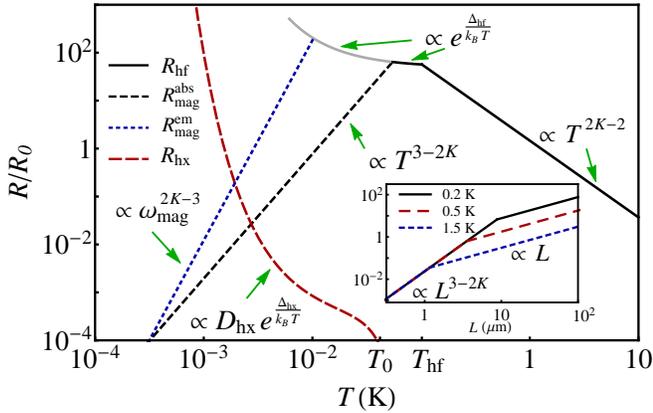}
\caption{
Temperature ($T$) dependence of the resistance $R$ for $L=25~\mu$m (so that $L>\xi_{\textrm{hf}}$) and $\lambda_{\textrm{mfp}}= 1~\mu$m. Above $T_{0}$, $R$ is given by Eqs.~(\ref{Eq:R_hf1}) and (\ref{Eq:R_hf2}) for $T>T_{\textrm{hf}} $ and $T<T_{\textrm{hf}}$, respectively. 
Below $T_{0}$, $R$ consists of three contributions due to: 
magnon emission $R_{\textrm{mag}}^{\textrm{em}}$, magnon absorption $R_{\textrm{mag}}^{\textrm{abs}}$, and nuclear order-assisted backscattering on impurities $R_{\textrm{hx}}$. The gray curve is the upper limit on $R_{\textrm{mag}}^{\textrm{em}}+R_{\textrm{mag}}^{\textrm{abs}}$ (see the text for explanations). The $T$ dependence of $\omega_{\textrm{mag}}$ is given by $ \omega_{\textrm{mag}} (T) \propto  T^{-(2-2K)} [ 1- (T/T_{0})^{3-2K}/2] $ for $T<T_{0}$.
Inset: Length ($L$) dependence of $R$ for various $T$.
}
\label{Fig:R_T}
\end{figure}

{\it Experimental consequences.}
To make specific predictions which can be examined in experiments, in Fig.~\ref{Fig:R_T} we summarize the temperature dependence of the edge resistance, as it follows from the presented analysis. Decreasing the temperature from well above $T_{0}$, the resistance first grows as a power law, which changes into an exponential at $T_\textrm{hf}$ (the black solid curve; additional possibilities were discussed above). 
The trend reverses at around $T_{0}$ (nuclear ordering temperature), resulting in a local maximum here.
Below $T_0$, the resistance is first mainly due to magnons. The magnon emission typically dominates the absorption, and the resistance decays as a power law (the blue curve). 
Finally, at even lower $T$ the resistance is dominated by nuclear order-assisted backscattering on impurities and grows exponentially (the red curve).

As the behavior for $L< \xi_{\textrm{hf}}$ is very similar (not shown), we conclude that the power-law increase at high $T$, the peak around $T_{0}$, and the exponential growth at $T\to 0$ are robust features of the nuclear spin-induced resistance of a 2DTI edge.
In addition, assuming that the value of the parameter $K$ is known for a given sample, one can verify the power-law dependences of the resistance on the voltage $V$ and the edge length $L$. 

The theoretically proposed backscattering mechanisms~\cite{Wu:2006,Strom:2010,Maciejko:2009,Tanaka:2011,Hattori:2011,Delplace:2012,Lunde:2012,Crepin:2012,Budich:2012,Lezmy:2012,Schmidt:2012,Vayrynen:2013,Altshuler:2013,DelMaestro:2013,Kainaris:2014,Geissler:2014,Vayrynen:2016}, including our work here, generally lead to different $V$, $L$, and $T$ dependence of the edge resistance. They can therefore be, in principle, discriminated experimentally. However, the majority of these mechanisms depend strongly on the Luttinger interaction parameter $K$, which is typically unknown in current experiments. A direct comparison of theories to experiments is then difficult, while the extraction of the value of $K$ is highly non-trivial for the very same reason \cite{Vayrynen:2016}. Specifically for our mechanism, it should be most relevant for measurements satisfying conditions of mesoscopic length, $L \gg 1$ $\mu$m, dilution fridge temperature, $T \ll 1$ K, and very strong interactions, $K\ll1$. Since a setup allowing for the investigations of the length dependence was realized recently in InAs/GaSb~\cite{Mueller:2017}, we believe that the experimental verification of this mechanism is feasible.
  
In conclusion, our most important finding is that, generally, the nuclear spins suppress the conductance of a long 2DTI edge to zero at very low temperatures. The scaling with exponentials or $V^{2K-2}$, $L^{3-2K}$, and $T^{2K-2}$ power laws, as summarized in Figs.~\ref{Fig:Rhf_V} and \ref{Fig:R_T}, allows to distinguish the nuclear spins from alternative mechanisms for the 2DTI edge resistance.

\acknowledgments

We thank M. R. Delbecq for helpful discussions. This work was supported financially by the JSPS KAKENHI (Grant No.~16H02204), the Swiss National Science Foundation, and the NCCR QSIT.

\bibliography{HLL}

\clearpage
\onecolumngrid

\bigskip 

\begin{center}
\large{\bf Supplemental Material to `Nuclear spin-induced localization of the edge states in two-dimensional topological insulators' \\}

\fontsize{10}{12}
Chen-Hsuan Hsu$^{1}$, Peter Stano$^{1,2}$, Jelena Klinovaja$^{1,3}$, and Daniel Loss$^{1,3}$\\
{\it
$^{1}$RIKEN Center for Emergent Matter Science (CEMS), Wako, Saitama 351-0198, Japan \\
$^{2}$Institute of Physics, Slovak Academy of Sciences, 845 11 Bratislava, Slovakia and\\
$^{3}$Department of Physics, University of Basel, Klingelbergstrasse 82, CH-4056 Basel, Switzerland
}
\end{center}

\twocolumngrid
\setcounter{equation}{0}
\setcounter{figure}{0}
\setcounter{table}{0}
\setcounter{page}{1}
\setcounter{NAT@ctr}{0}   
\makeatletter
\renewcommand{\theequation}{S\arabic{equation}}
\renewcommand{\thefigure}{S\arabic{figure}}
\renewcommand{\bibnumfmt}[1]{[S#1]}
\renewcommand{\citenumfont}[1]{S#1}

\section{hyperfine interaction}

In this supplemental section we estimate the order of magnitudes of the hyperfine interaction arising from the Fermi contact, dipolar, and orbital contributions per nucleus. We start by writing the vector potential generated by a nuclear spin ${\bf I}$,
\begin{eqnarray}
{\bf A}_n &=& \frac{\mu_0} {4\pi} \frac{ \hbar \gamma_{n} {\bf I} \times {\bf r}} {r^3},
\end{eqnarray}
with the magnetic moment $\hbar \gamma_{n} {\bf I}$ carried by the nuclear spin and the gyromagnetic ratio $\gamma_n$. Here $r = |{\bf r}|$ is the distance between the electron and nucleus with ${\bf r} = (x,y,z)$. The magnetic field $\triangledown \times {\bf A}_n $ is then felt by an electron with spin ${\bf S}$, leading to the Fermi contact and dipolar contributions~\citesupp{Okvatovity:2016_S,Lunde:2013_S}, 
\begin{eqnarray}
H_{\textrm{hf}}^{\textrm{Fc}} &=& -\frac{\mu_0} {4\pi} \frac{8\pi} {3} \hbar \gamma_{n} g_{e} \mu_{B} {\bf S} \cdot {\bf I} \delta({\bf r}), \\
H_{\textrm{hf}}^{\textrm{dip}} &=& \frac{\mu_0} {4\pi} \hbar \gamma_{n} g_{e} \mu_{B} {\bf I}  \cdot
 \left[ \frac{  {\bf S} - 3 \hat{r} ({\bf S} \cdot  \hat{r} )  } {r^3} \right],
\end{eqnarray}
with $g_{e}$ and $\mu_{B}$ being the electron gyromagnetic constant and Bohr magneton, respectively.
Following Ref.~\citesupp{Okvatovity:2016_S}, we obtain the orbital contribution with the Peierls substitution of ${\bf A}_n$ into the electron kinetic energy $H_{\textrm{kin}}$. For the edge states of two-dimensional topological insulators (2DTIs), we have $ H_{\textrm{kin}}({\bf p}) = v_F \sigma^z p^x$ with the Pauli matrix $\sigma^z$ and momentum $p^x = -i\hbar \partial_x $, from which we get
\begin{eqnarray} 
 H_{\textrm{hf}}^{\textrm{orb}} &=& H_{\textrm{kin}} ({\bf p} - e {\bf A}_n) - H_{\textrm{kin}} ({\bf p}) \nonumber\\
&=&  \frac{\mu_0} {4\pi} \hbar \gamma_{n} e v_F  \sigma^z  \frac{ ( y I^z - z I^y ) } {r^3}. 
\label{Eq:H_orb_2DTI}
\end{eqnarray}
This term, being proportional to $\sigma^z$, does not lead to the electron spin flip, and is not relevant to our analysis in the disordered phase, i.e., it does not enter the backscattering strength $D_{\textrm{hf}}$ in the main text. It does not contribute to the RKKY coupling, either. Nonetheless, we estimate its magnitude below, as well as the Fermi contact and dipolar contributions. 

To proceed, we estimate the energy scales of the following matrix elements~\citesupp{Okvatovity:2016_S}, 
\begin{subequations} 
\label{Eq:M_hf}
\begin{eqnarray}
\left < \Psi \left| H_{\textrm{hf}}^{\textrm{Fc}} \right| \Psi'\right> & = & -\frac{2\mu_0} {3} \hbar \gamma_{n} g_{e} \mu_{B} \left < \Psi \left| {\bf S} \cdot {\bf I} \delta({\bf r}) \right| \Psi'\right> , \label{Eq:M_fc} \\
\left < \Psi \left| H_{\textrm{hf}}^{\textrm{dip}} \right| \Psi'\right>  &=& \frac{\mu_0} {4\pi} \hbar \gamma_{n} g_{e} \mu_{B} 
\left < \Psi \left| {\bf I}  \cdot  \left[ \frac{  {\bf S} - 3 \hat{r} ({\bf S} \cdot  \hat{r} )  } {r^3} \right] \right| \Psi'\right> ,\nonumber \\ \label{Eq:M_dip} \\
\left < \Psi \left| H_{\textrm{hf}}^{\textrm{orb}} \right| \Psi'\right> 
&=& \frac{\mu_0} {4\pi} \hbar \gamma_{n} e v_F   \left< \Psi \left|  \sigma^z  \frac{ ( y I^z - z I^y ) } {r^3} \right| \Psi'\right>, \label{Eq:M_orb}
\end{eqnarray}
\end{subequations}
where we choose the initial $\left| \Psi' \right>$ and final $\left| \Psi \right>$ states to be the edge states with the opposite (same) velocities for the spin-flip $S^{x,y}$ (spin-conserving $S^{z}$) terms. 
To proceed, we express the edge states as the product of the Bloch amplitude $u_{\textrm{B}}$ and the envelope function ($\left| \Psi_{R/L} ({\bf r})\right>$ for the right/left state),
\begin{eqnarray}
\left| \Psi \right>,~\left|\Psi'\right> &=& u_{\textrm{B}}({\bf r}) \left| \Psi_{R/L} ({\bf r}) \right>,
\end{eqnarray}
where the Bloch amplitude $u_{\textrm{B}} ({\bf r})$ satisfies $\int d{\bf r} \; |u_{\textrm{B}} ({\bf r})|^2  f({\bf r}) \approx \int d{\bf r} \; f({\bf r})$ for functions $f({\bf r})$ that are smooth over the atomic scale $a_0$. We assume that $\left| \Psi_{R/L} ({\bf r})\right>$ can be factorized into the longitudinal and transverse parts,
\begin{eqnarray}
\left| \Psi_{R/L} ( {\bf r}) \right> &=& C_{\parallel} e^{\pm i k_{F} x} \left| \downarrow/\uparrow \right>  \otimes \left| \Psi_{\perp} (y,z) \right>,
\end{eqnarray}
where the longitudinal part is written as the product of the spatial part (with the normalization factor $C_{\parallel} =1/\sqrt{L} $) and the spin state $\left| \sigma \right>$, and the transverse part fulfills $\left< \Psi_{\perp} (y,z)\right. \left| \Psi_{\perp} (y,z) \right> =1$. We now estimate the matrix elements in Eqs.~(\ref{Eq:M_hf}).

\subsection{I. Fermi contact contribution}
We start with the Fermi contact contribution. Here we compute the matrix element of the spin-flip $S^{+}\equiv S^{x} + i S^{y}$ term (the spin-conserving $S^{z}$ term should give similar results), 
\begin{eqnarray}
E_{\textrm{Fc}} &\equiv&  \frac{2\mu_0}{3} \hbar \gamma_{n} g_{e} \mu_{B}  \left < \Psi \left|  S^{+} \delta({\bf r}) \right| \Psi'\right>  \nonumber \\
 &=&  \frac{2\mu_0}{3} \hbar \gamma_{n} g_{e} \mu_{B} \frac{1}{L w^2} \left|u_{\textrm{B}}( {\bf r}=0) \right|^2, 
\label{Eq:E_fc}
\end{eqnarray}
where $w$ is the transverse length scale, and the values of $\eta \equiv \left|u_{\textrm{B}}( {\bf r}=0)\right|^2 $ for the relevant nuclei were estimated in semiconductor systems~\citesupp{Gueron:1964_S,Paget:1977_S,Schliemann:2003_S}. Although Refs.~\citesupp{Gueron:1964_S,Paget:1977_S,Schliemann:2003_S} are not about 2DTI materials (e.g. InAs and GaSb compounds), we expect that the $\eta$ value here to be of the same order of the magnitudes. For the Hg and Te nuclei in HgTe, however, the value of $\eta$ may be smaller due to the larger principal quantum number of the outermost electrons, as discussed in Ref.~\citesupp{Lunde:2013_S}.

\subsection{II. Dipolar contribution}
We now turn to the dipolar contribution to the hyperfine interaction. Because the spatial dependence in Eq.~(\ref{Eq:M_dip}) is smooth over the atomic scale, the details of $u_{\textrm{B}}({\bf r})$ do not play a role. Since we are only interested in the overall scale, we simplify the spatial dependence as $r^{-3}$, and estimate the spin-flip $S^{+}$ term in Eq.~(\ref{Eq:M_dip}), 
\begin{eqnarray}
E_{\textrm{dip}} 
&\equiv& \frac{\mu_0} {4\pi} \hbar \gamma_{n} g_{e} \mu_{B} C_{\parallel}^2 \int_{-\infty}^{\infty} dx \; e^{ 2i k_{F} x} \left < \Psi_{\perp} (y,z) \left| \frac{ 1 } {r^3} \right| \Psi_{\perp} (y,z) \right>.  \nonumber\\
\end{eqnarray}
Assuming that the integrals over $y$ and $z$ coordinates lead to
\begin{eqnarray}
\left < \Psi_{\perp} (y,z) \left| \frac{ 1 }{r^3} \right| \Psi_{\perp} (y,z) \right> &=& \frac{1}  { (x^2 + w^2)^{3/2}},
\end{eqnarray}
and performing the remaining $x$ integral, we get 
\begin{eqnarray}
E_{\textrm{dip}}  &\sim& \frac{\mu_0} {4\pi} \hbar \gamma_{n} g_{e} \mu_{B} \frac{0.1}{L w^2},
\label{Eq:E_dip}
\end{eqnarray}
where the factor $0.1$ comes from the suppression due to the oscillatory integrand. Comparing with the Fermi contact contribution Eq.~(\ref{Eq:E_fc}) gives
\begin{eqnarray}
\frac{E_{\textrm{dip}} } {E_{\textrm{Fc}}} &\sim& \frac{3}{80\pi\eta} \sim O(10^{-2}) \times \frac{1}{\eta},
\end{eqnarray}
which means the dipolar contribution is at least two orders smaller than the Fermi contact contribution as long as $\eta>1$, even though we do not know the exact $\eta$ value in InAs/GaSb.   
Therefore, the dipolar contribution to the hyperfine interaction is much weaker than the Fermi contact contribution.

\subsection{III. Orbital contribution}
We now turn to the orbital contribution. Even though the orbital contribution does not cause spin flip, we still estimate its magnitude for the sake of completeness. Again, we use the fact that $y/r^3$ and $z/r^3$ are smooth on the atomic scale, so the details of $u_{\textrm{B}}( {\bf r})$ can be neglected. The matrix element Eq.~(\ref{Eq:M_orb}) can be written as
\begin{eqnarray}
 \frac{\mu_0} {4\pi} \hbar \gamma_{n} e v_F  C_{\parallel}^2 \int_{-\infty}^{\infty} dx \; \left < \Psi_{\perp} (y,z) \left|  \frac{ ( y I^z - z I^y ) } {r^3}  \right| \Psi_{\perp} (y,z) \right>. \nonumber \\
\end{eqnarray}
To proceed, we assume the integral over the transverse part can be written as,
\begin{subequations}
\begin{eqnarray}
\left < \Psi_{\perp} (y,z) \left|   \frac{  y  } {r^3}  \right| \Psi_{\perp} (y,z) \right> &=& \frac{w_{y}}  { (x^2 + w^2)^{3/2}} ,\\
\left < \Psi_{\perp} (y,z) \left|   \frac{  z  } {r^3}  \right| \Psi_{\perp} (y,z) \right> &=& \frac{w_{z}}  { (x^2 + w^2)^{3/2}},
\end{eqnarray}
\end{subequations}
with $w^2 =w_y^2 + w_z^2 $. Performing the remaining $x$ integral, we obtain the magnitude of Eq.~(\ref{Eq:M_orb}),
\begin{eqnarray}
 E_{\textrm{orb}} &\sim&  \frac{\mu_0} {4\pi} \hbar \gamma_{n} e v_F  \frac{1}{L w}. 
\end{eqnarray}
Comparing with the Fermi contact contribution Eq.~(\ref{Eq:E_fc}) gives
\begin{eqnarray}
\frac{E_{\textrm{orb}} }{E_{\textrm{Fc}}} &\sim& \frac{3}{8\pi} \frac{e v_F w}{g_e \mu_B \eta} \sim O(10^{-1}) \times \frac{1}{\eta},
\end{eqnarray}
where we have used the parameters, $\mu_0=4\pi \times 10^{-7}$~Vs/Am, $g_e=2$, $\mu_B= 5.8 \times 10^{-5}$~eV/T, $v_{F}=4.6\times 10^4~$m/s, and $w = 10~$nm. As mentioned above, even though we do not have the exact $\eta$ value, we find the above ratio to be much smaller than 1 for any $\eta>1$.  
As a summary, both the dipolar and orbital contributions to the hyperfine interaction are much weaker than the Fermi contact contribution.

\subsection{IV. Hyperfine coupling}
We note that the energy scale $E_{\textrm{Fc}}$ in Eq.~(\ref{Eq:E_fc}) is not the hyperfine coupling $A_0$ in the main text. Since we define $A_0$ such that $H_{\textrm{hf}} \sim (A_0/\rho_{\textrm{nuc}}) \rho_{e} {\bf S} \cdot {\bf I} $ with the nuclear and electron densities, $\rho_{\textrm{nuc}}=8/a_0^3$ and $\rho_{e} \sim 1/(Lw^2)$, respectively, we have
 \begin{eqnarray}
A_0 &\sim & \frac{16\mu_0}{3} \hbar \gamma_{n} g_{e} \mu_{B} \frac{\eta}{a_{0}^3}. 
\label{Eq:A_0}
\end{eqnarray}
In Refs.~\citesupp{Gueron:1964_S,Paget:1977_S,Schliemann:2003_S}, the $\eta$ values for the relevant nuclei are given by $\eta_{\textrm{In}} = 6.3 \times 10^3$, $\eta_{\textrm{Sb}} = 1.1 \times 10^4$, $\eta_{\textrm{Ga}} = 2.7 \times 10^3$, and $\eta_{\textrm{As}} = 4.5 \times 10^3$. If we take the average value, $\eta=6.1 \times 10^3$, for our estimation, along with the parameters $\hbar \gamma_{n}=6 \times 10^{-8}~$eV/T, $a_0=6.1~${\AA}, $g_e=2$, and $\mu_B= 5.8 \times 10^{-5}$~eV/T, we obtain $A_{0}\sim O(100~\mu\textrm{eV})$. Since, however, the edge states are mixtures of $s$- and $p$-orbital states, the actual contribution from the $s$-orbital state may be somewhat smaller. We note that Ref.~\citesupp{Lunde:2013_S} investigated the hyperfine interaction in HgTe 2DTIs, and found that, for the spin-flip terms, the contribution from the Fermi contact dominates over the other contributions.
We also note that generalizing our model with an anisotropic hyperfine coupling ($A_x \neq A_y$) may modify the backscattering strength, e.g. by replacing $A_0^2 \rightarrow (A_x^2+A_y^2)/2$, but does not lead to a qualitative difference. As a result, we keep only the Fermi contact contribution in the main text for our analysis.

\section{Dipole-dipole interaction of nuclei}

In this supplemental section we discuss the effects of the dipole-dipole interaction between the nuclear spins on the electron backscattering. The nuclear dipolar interaction is much weaker than the electron-nuclear hyperfine interaction~\citesupp{Paget:1977_S}. It is nevertheless an important ingredient for the dissipation of the nuclear spin polarization. 
The (dynamical) polarization would otherwise accumulate during spin-flip backscattering processes~\citesupp{Lunde:2012_S,Kornich:2015_S} and therefore prevent subsequent backscattering events. Even though the nuclear dipolar interaction is not explicitly written in our Hamiltonian Eqs.~(\ref{eq:H_LL})--(\ref{Eq:H_hf}), we implicitly include it in our analysis by considering that the nuclear subsystem is in its (thermal) ground state, whether ordered or not. The dipole-dipole nuclear spin diffusion is the mechanism for the nuclear subsystem to return to this ground state upon excitations by current.
We note that this makes our theory different from, e.g. backscattering on a single magnetic impurity or a spin bath where such a dissipation channel is absent, and the backscattering is trivially shut down once these magnetic impurities become polarized.

\section{RKKY interaction}

In this supplemental section we discuss the RKKY interaction mediated by a helical Tomonaga-Luttinger liquid. Similar to nonhelical systems~\citesupp{Simon:2008_S,Braunecker:2009a_S,Braunecker:2009b_S,Meng:2014a_S,Hsu:2015_S}, here we integrate out the electronic degrees of freedom in the hyperfine interaction to obtain the RKKY interaction between the localized nuclear spins. The interaction strength is proportional to the electronic spin susceptibility, and can be calculated along the line of Ref.~\citesupp{Giamarchi:2003_S}. The RKKY strength develops a dip at $q=2k_{F}$, with the magnitude,
\begin{eqnarray}
 J^{x}_{2k_{F}} &\approx& - \frac{  \sin (\pi K) }{8\pi^2} \frac{K A_{0}^2}{ \Delta }
\left(\frac{\lambda_{T}}{2\pi a} \right)^{2-2K} 
\left| \frac{ \Gamma\left( 1-K\right) \Gamma\left( \frac{K}{2} \right) }
{\Gamma\left(\frac{2-K}{2} \right)}\right|^{2}, \nonumber \\
\label{Eq:JRKKY}
\end{eqnarray}
with the Gamma function $\Gamma(x)$. Equation~(\ref{Eq:JRKKY}) is then used in the calculation of the magnon energy $\hbar \omega_{\textrm{mag}}$ and the resistance due to the magnon emission $R_{\textrm{mag}}^{\textrm{em}}$ [Eq.~(\ref{Eq:R_mag})] in the main text.

\section{Interplay of the electronic and nuclear subsystems}

\begin{figure}[t]
\centering
\includegraphics[width=0.5\linewidth]{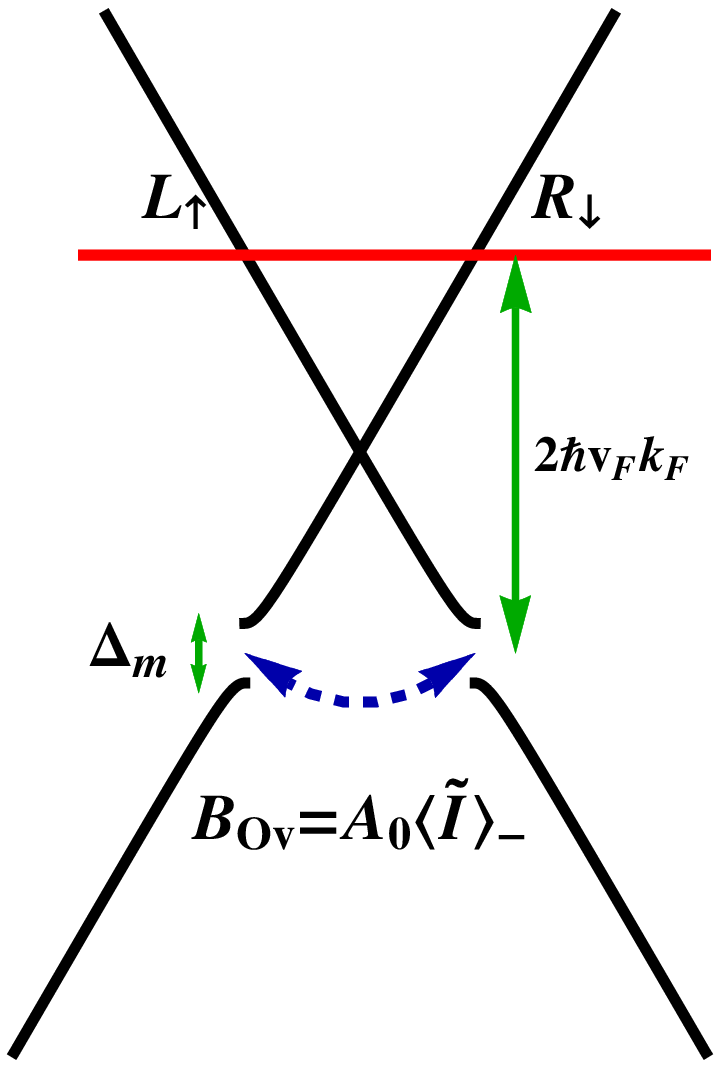}
\caption{Backscattering due to the Overhauser field arising from the nuclear spin order with the ground state value $\left< \tilde{I}\right>_{-}$. A gap $\Delta_{\textrm{m}}$ opens below the Fermi surface.}
\label{Fig:backscattering}
\end{figure}

In this supplemental section we comment on two important features of the interplay of the electronic and nuclear subsystems.
First, in the ordered phase, the Overhauser field arising from the ordered nuclear spins induces an electronic gap $\Delta_{\textrm{m}}$ below the Fermi surface, as shown in Fig.~\ref{Fig:backscattering}. This gap may provide for additional experimental signatures of the nuclear spin order. Due to the separation of the time scales of the electron and nuclear subsystems, one can rapidly change the chemical potential via a gate voltage, while the spatial modulation of the ordered nuclear spins (thus the Overhauser field and the position of the gap) remains intact. Therefore, by sweeping the chemical potential across the gap, experimental signatures can be sought via techniques that were employed to detect the effects of the nuclear spins in semiconducting systems, such as transport, optical, and NMR measurements~\citesupp{Dobers:1988_S,Smet:2002_S,Chekhovich:2013_S,Tiemann:2014_S}.

Second, a finite gap $\Delta_{\textrm{hx}}$ due to the nuclear order-assisted backscattering on impurities reduces the RKKY coupling, which we did not take into account in the main text. This mechanism imparts a negative feedback onto the effects of the nuclear spin order on the resistance. 
With the exact solution being beyond the scope of this work, we only remark that we expect our results to remain qualitatively valid as long as $\Delta_{\textrm{hx}} \ll \epsilon_{F}$, which preserves the sharp RKKY dip around momentum $2k_F$, albeit with a reduced height~\citesupp{Klinovaja:2013b_S,Meng:2014a_S,Hsu:2015_S}.

\bibliographystylesupp{apsrev4-1}
\bibliographysupp{HLLsupp}

\end{document}